\documentclass[aps,manuscript,showpacs,psfig]{revtex4}
\usepackage{graphicx}
\usepackage{psfig}

\begin{document}

\titlepage
\setcounter{page}{1}
\title{Constraints of the Dynamic Equations and Lagrangian required by Superposition of  Field}
\author{ X. Sun$^a$, Z. Yang$^{b,c}$\\
$^a$Institute of High Energy Physics, Beijing 100039, China\\
$^a$Graduate University of the Chinese Academy of Sciences, Beijing 100039, China\\
$^b$Department of Engineering Physics, Tsinghua University,
Beijing 100084, China\\
$^c$Center of High Energy Physics, Tsinghua University, Beijing
100084, China}

\begin{abstract}
A general form of the dynamical equations of field is obtained on
the requirement  this field is a superposable one; hence the
constraint on the forms of the Lagrangians is acquired. It shows
this requirement requires the continuous transformation group of the
Lagrangians of field to be compact, and that All Lagrangians of
elementary particles, such as leptons, quarks, photons and gluons,
satisfy this requirement.The result of regarding this character as a
general property of physical field is discussed.
\end{abstract}
 \maketitle
%\narrowtext
%\twocolumn
\newpage

\section{Introduction}
 Superposition principle of quantum states and
Schr\"odinger equation are two postulations of non-relativistic
quantum mechanics. Then some other equations, such as Klein-Gordon
equation and Dirac equation, are founded in other fields of
quantum theory. As the quantum field theory is used in many
aspects, many kinds of Lagrangians are born to solve different
kinds of problems. Restrictions to these Lagrangians are generally
Lorentz covariance, gauge invariance and renormalizibility. On the
other hand, superposition principle of quantum states has been
fully discussed\cite{Dur,Enk,JKL01,Wang,JK,Ralph} and is examined
by quite many experiments\cite{att,rose,Ralls,quist,berg}. It has
been accepted as a basic rule of microscopic world. In some basic
processes such as creation and annihilation of particles, quantum
field theory is tested to be successful under many conditions. As
the wave function in quantum mechanics can also be regarded as a
kind of field,we start from a superposable field,then get the
universal form of the dynamical equations and Lagrangians of the
field.

In fact, the property of superposition may lead to some additional
restrictions on the form of equations of motion, and therefore on
that of Lagrangians of dynamic systems.

In the present paper, we try to argue the constraints of
Lagrangians of field caused by that the field can be superposed.

The outline of this paper is as follows. In section II, we first
obtain the general form of the equations of motion starting from a
field with the property of superposition. The constraints on
Lagrangians are then argued in section III. All the successful
Lagrangians, such as scalar field, Dirac field, QED and QCD,and
some of interesting fields are discussed in  section III. A brief
conclusion is made in section IV.

\section{Constraints on Equation of motion Caused by the property of superposition}

In quantum theory, the wave function, $\psi(\vec{x},t)$, is the
probability amplitude whose absolute square indicates the
probability to find a particle in position $\vec{x}$ at time $t$.
The evolution operator $\hat{U}(\vec x_2,t_2;\vec x_1,t_1)$ means
the probability amplitude for a particle, originally in position
$\vec x_1$ at time $t_1$, to be find in position $\vec x_2$ at
time $t_2$. The evolution operator can be expressed in form of
integral operator of propagator. After operating on $\psi(\vec
x_1, t_1)$ from left-hand side, it gives wave function $\psi(\vec
x_2, t_2)$, i.e.
\begin{eqnarray}
 \label{eq1}
  & &\hat{U}(\vec x_2,t_2;\vec x_1,t_1)\psi(\vec x_1,t_1)\nonumber\\
  &=&\int d \vec x_1 K(\vec x_2,t_2;\vec x_1,t_1)\psi(\vec x_1,t_1)=\psi(\vec x_2,t_2).
\end{eqnarray}
Now we extract the physical meaning of $\psi(\vec{x},t)$,treating
it as a field of time-space coordinate.$\hat{U}(\vec x_2,t_2;\vec
x_1,t_1)$ still is a functional of $\psi(\vec{x},t)$ as before. We
still call $\psi(\vec{x},t)$,$\hat{U}(\vec x_2,t_2;\vec x_1,t_1)$
as wave function, evolution operator respectively for the
understanding of our deduction .$\psi(\vec{x},t)$ is an
representation of fields with the property of superposition
without any physical meaning.the operation between
$\psi(\vec{x},t)$ and $\hat{U}(\vec x_2,t_2;\vec x_1,t_1)$ is just
like (\ref{eq1}).

the property of superposition of $\psi(\vec{x},t)$

A dynamic equation of $\hat{U}(\vec x_2,t_2;\vec x_1,t_1)$  can be
expressed by
\begin{eqnarray}
 \label{eq2}
 \hat{F}\hat{U}(\vec x_2,t_2;\vec x_1,t_1)=0.
\end{eqnarray}

Let both sides of equation (\ref{eq2}) operate on $\psi(\vec
x_1,t_1)$ from left-hand side. It is easy to show that the wave
function satisfies the same dynamic equation to that of the
evolution operator.
\begin{eqnarray}
 \label{eq3}
  \hat{F}\hat{U}(\vec x_2,t_2;\vec x_1,t_1)\psi(\vec x_1,t_1)=\hat{F}\psi(\vec x_2,t_2)=0.
\end{eqnarray}

Final state is the superposition of all the transitional states.
Mathematically, the property of superposition of $\psi(\vec{x},t)$
can be expressed by
\begin{eqnarray}
 \label{eq4}
  & & K(\vec x_3,t_3;\vec x_1,t_1)\nonumber\\
  &=&\int d x_2 K(\vec x_3,t_3;\vec x_2,t_2)K(\vec x_2,t_2;\vec x_1,t_1).
\end{eqnarray}

For simplicity, we set $t_1=0,\ t_2=t$ and $t_3=t+\Delta t$. The
expansion of $K(\vec x_3,t+\Delta t;\vec x_1,0)$ to the first
order approximation can be written as
\begin{eqnarray}
 \label{eq5}
  &       &K(\vec x_3,t+\Delta t;\vec x_1,0)\nonumber\\
  &\approx&K(\vec x_3,t;\vec x_1,0)+\frac{\partial K(\vec x_3,t;\vec x_1,0) }{\partial t}\Delta t.
\end{eqnarray}
It is trivial to get
\begin{eqnarray}
 \label{eq6}
  \frac{\partial K(\vec x_3,t;\vec x_1)}{\partial t}=\frac{K(\vec x_3,t+\Delta t;\vec x_1)- K(\vec x_3,t;\vec x_1)}{\Delta t},
\end{eqnarray}
where the index of initial time $t_1=0$ is omitted.

Now setting $\vec x_2=\vec x_3-\Delta \vec x$, the integrand in
equation (\ref{eq4}) can be expanded as
\begin{eqnarray}
 \label{eq7}
  K(\vec x_3,t_3;\vec x_2,t_2)K(\vec x_2,t_2;\vec x_1)=\sum_{n=0}^\infty \frac{(-\Delta \vec x)^n}{n!}F_n,
\end{eqnarray}
with
\begin{eqnarray}
 F_n\equiv\frac{\partial^n[K(\vec x_3+\Delta \vec x,t+\Delta t;\vec x_3,t)K(\vec x_3,t;\vec x_1)]}
               {\partial x_3^n}.\nonumber
\end{eqnarray}

Substitute equation (\ref{eq4}) and (\ref{eq7}) into (\ref{eq6}),
and it follows that
\begin{eqnarray}
 \label{eq8}
 & &\frac{\partial}{\partial t}K(\vec x_3,t;\vec x_1)\nonumber\\
 &=&\frac{1}{\Delta t}\left[\int d(\Delta \vec x)
                            \sum_{n=0}^\infty\frac{(-\Delta \vec x)^n}{n!}F_n-K(\vec x_3,t;\vec x_1)\right]\nonumber\\
 &=&\frac{1}{\Delta t}\left[\sum_{n=0}^\infty\frac{1}{n!}\frac{\partial^n}{\partial \vec x_3^n}
                            G(\vec x_3,n)K(\vec x_3,t;\vec x_1)-K(\vec x_3,t;\vec
                            x_1)\right],\nonumber\\
\end{eqnarray}
where $G(\vec x_3,n)$, with the time index omitted, is defined as
\begin{eqnarray}
 G(x_3,n)\equiv\int d(\Delta\vec x)(-\Delta\vec x)^nK(\vec x_3+\Delta \vec x,t+\Delta t;\vec
 x_3,t).\nonumber
\end{eqnarray}

Note that for $n=0$, $G(\vec x_3,n)$ is nothing but the total
probability of finding a particle in the whole space with initial
space-time coordinates $(\vec x_3, t)$. This indicates that $\int
d(\Delta x) K(x_3+\Delta x,t+\Delta t;x_3,t)=1$. Therefore,
equation (\ref{eq8}) can be reduced to
\begin{eqnarray}
 \label{eq9}
  \frac{\partial}{\partial t}K(\vec x_3,t;\vec x_1)
 =\sum_{n=1}^\infty\frac{\partial^n}{\partial \vec x_3^n} S^1_nK(\vec x_3,t;\vec x_1),
\end{eqnarray}
with
\begin{eqnarray}
\label{eq10}
 S^1_n=\frac{1}{n!\Delta t}G(\vec x_3,n),\ \ \ \  n=1,2,3,\cdots
\end{eqnarray}

Operate both sides of equation (\ref{eq9}) on the wave function
$\psi(\vec x_1,t_1)$ from left side, and integrate over $\vec x_1$
we acquire, with substituting $\vec x$ for $\vec x_3$,
\begin{eqnarray}
 \label{eq11}
  \frac{\partial}{\partial t}\psi(\vec x,t)=\sum_{n=1}^\infty\frac{\partial^n }{\partial \vec x^n} S^1_n\psi(\vec x,t).
\end{eqnarray}

When the second order term in (\ref{eq5}) is considered, we can
get, by following similar deductions as above,
\begin{eqnarray}
\label{eq13}
 \frac{\partial^2 }{\partial t^2}\psi(\vec x,t)=\sum_{n=1}^\infty\frac{\partial^n}{\partial \vec x^n}S^2_n\psi(\vec
 x,t).
\end{eqnarray}
For general, it can be deduced that
\begin{eqnarray}
\label{eq15}
 \frac{\partial^m }{\partial t^m}\psi(\vec x,t)=\sum_{n=1}^\infty\frac{\partial^n }{\partial
 \vec x^n}S^m_n\psi(\vec x,t),
\end{eqnarray}
with
\begin{eqnarray}
 \label{eq14}
 S^m_n=\frac{m!}{n!(\Delta t)^m}G(\vec x_3,n),\ \ \ \  n,m=1,2,3,\cdots
\end{eqnarray}

Now we get the general form of dynamic equation, due to the
requirement of the property of superposition of $\psi(\vec{x},t)$,
defined as equation (\ref{eq15}) or the linear nestification of it
for different $m$.

\section{Constraints on Lagrangians by the property of superposition}

Restriction on dynamic equation will limit the forms of
Lagrangians, ${\cal L}$, of the system, since Euler equation
\begin{eqnarray}
\label{eq16}\partial_\mu\frac{\partial{\cal
L}}{\partial(\partial_\mu \psi)}-\frac{\partial{\cal L}}{\partial
\psi}=0
\end{eqnarray}
gives the dynamic equations for a system described by $\cal L$.
After quantization, $\psi$ becomes field operator instead of wave
function. Considering the fact that the Lagrangian contains only
coordinates, $x_\mu$, and one-order differential to coordinates
$\frac{\partial}{\partial x_\mu}$, but not any other higher order
differential terms, the general dynamic equation derived from the
property of superposition, equation (\ref{eq15}), gives a general
form of Lagrangian
\begin{eqnarray}
 \label{eq17}\
 {\cal L}=\sum_k C_k\psi^{i}(\partial_{\mu}\psi)^{j},\ \ \  i,j=0,1,2.
\end{eqnarray}
where $C_k$ is coefficient of the $k^{th}$ item, and $i,j$ can be
selected from $0, 1, 2$, according to other requirements, such as
dimensional requirement and Lorentz invariance. But it is
forbidden that both $i$ and $j$ are selected to be $2$. Surely
many other constraints should be concluded. For example, Lorentz
invariance require the Lorentz indexes should be contracted. the
coefficient $C_k$ may contain other fields, if existing, which
gives the interaction between fields.

Do all the Lagrangians describing elementary fields, which have
been used successfully, meet the requirement deduced from the
property of superposition? We first examine the Lagrangians of
scalar field, Dirac field and electromagnetic field, which
correspond to fields with spin $s=0, \frac{1}{2}, 1$,
respectively. The Lagrangians with no interactions of these fields
are as follows,
\begin{eqnarray}
 \label{eq18}
 {\cal L}_s &=& \frac{1}{2}\partial^\mu\varphi\partial_\mu\varphi-\frac{1}{2}m^2\varphi^2\nonumber\\
 {\cal L}_D&=&i\bar{\psi}\gamma^\mu\partial_\mu\psi-m\bar{\psi}\psi\nonumber\\
 {\cal L}_{em}&=&-\frac{1}{4}{\cal F}_{\mu\nu}{\cal F}^{\mu\nu}\nonumber\\
              &=&-\frac{1}{2}(\partial_\mu A_\nu\partial^\mu A^\nu-\partial_\mu A_\nu\partial^\nu A^\mu)
\end{eqnarray}

The kinetic term of scalar field can be expressed in the form of
equation (\ref{eq17}) as $C_1=1/2$, $i=0$, and $j=2$. While the
mass term can be expressed as $C_2=\frac{1}{2}$, $i=2$ and $j=0$.
For Dirac field, the kinetic term, in the form of equation
(\ref{eq17}), can be expressed as $C_1=i\bar\psi\gamma^\mu$, $i=0$
and $j=1$. And the mass term is $C_2=m\bar\psi$, $i=1$ and $j=0$.
For electromagnetic field, it corresponds to $i=0$ and $j=2$. So
all the free fields of these elementary particles match the
requirement by the property of superposition.

Let us then consider the interaction term in quantum
electrodynamics(QED). The interaction between photons and
electrons is
\begin{eqnarray}
 \label{eq19}
 {\cal L}_i=-e\bar{\psi}\gamma^\mu\psi A_\mu.
\end{eqnarray}
According to equation (\ref{eq17}), $i=1$, $j=0$ and
$C_k=-e\bar\psi\gamma^\mu A_\mu$ are chosen for electron field
$\psi$. We surely can also choose $i=1$, $j=0$ and
$C_k=-e\bar\psi\gamma^\mu\psi$, if we consider it as a term for
photon field, $A_\mu$.

Now it comes to the Lagrangian of quantum chromodynamics(QCD). The
classical part of QCD Lagrangian, which is invariant under local
$SU(N_c)$ (with $N_c=3$ for QCD) gauge transformation, is defined
as
\begin{eqnarray}
 \label{eq_QCDclassical}
  {\cal L} &=& \sum_f\bar\psi_f[i\gamma^\mu D_\mu-m_f]\psi_f +{\cal  L}_g,\\
 \label{eq20}
  {\cal L}_g &=& -\frac{1}{4}{\cal F}_{\mu\nu}^a{\cal F}^{\mu\nu,a},
\end{eqnarray}
where $D_\mu$ is the covariant derivative, and $F_{\mu\nu}^a$, the
gluon-field strength. Similarly, it is easy to find that the first
term in equation (\ref{eq_QCDclassical}) satisfies the requirement
of the property of superposition. With the definition of
$F_{\mu\nu}^a$, the pure non-Abelian gauge field part, ${\cal
L}_g$, can be written as
\begin{eqnarray}
 \label{eq20}
  {\cal L}_g &=&-\frac{1}{4}(\partial _\mu A^a_\nu-\partial _\nu A^a_\mu)^2
                -g(\partial_\mu A^a_\nu)f^{abc}A^{\mu,b} A^{\nu,c}\nonumber\\
             & &-\frac{1}{4}g^2 f^{abc}f^{ab'c'}A_\mu^b A_\nu^c A^{\mu,b'}A^{\nu,c'},
\end{eqnarray}
where $f^{abc}$ is the structure constant of gauge group and $g$
the coupling constant. It is obvious that the first two terms,
which correspond to the kinetic part and 3-gluon interaction part
respectively, agree with the form of equation (\ref{eq17}).
However, the last term, i.e., the 4-gluon interaction term in QCD,
seems to break the rules described in equation (\ref{eq17}), since
it is a quartic term of non-Abelian gauge field and may brings a
term of $A^4$ when $b=b'$ and $c=c'$. But this term vanishes if
the coefficient of this term is zero for $b=b'$ and $c=c'$. To do
this, the coefficient must be antisymmetric, which means
$f^{abc}=\varepsilon_{abc}^{ab'c'}f^{ab'c'}$. This shows that the
property of superposition of field requires the continuous group
of transformation be a compact Lie group. Fortunately, this is
satisfied by the present theories where $SU(N)$ groups are used.
It is not hard to verify that the gauge-fixing term and
Faddeev-Popov term, which are introduced to quantize the classical
QCD, also meet the constraint of the property of superposition.
Upper fields are successful and tested in both theory and
experiment.So all the successful and tested fields of elementary
particle are superposable by now. Is it incidental?We don't think
so.We promote it as a basic rule that physical fields of
elementary particle are superposable.

A challenge of this opinion comes from $\varphi^4$ field.It is one
kind of interaction that is obvious in contradiction with
(\ref{eq17}) . $\varphi^4$ interaction is widely used as an
example in textbooks of quantum field theory, because it is one of
the simplest interaction and is renormalizable. But this
interaction violates the property of superposition of field
apparently, since it corresponds to $i=4$ and $j=0$ if expressed
in the form of equation (\ref{eq17}). In fact, no physical
particle with $\varphi^4$ interaction is found up to now.

It should be emphasized that the term introduced to a Lagrangian,
in order to manifest spontaneous symmetry breaking, may violate
the property of superposition of field. For example, the
introduction of Higgs field, which plays an important role in
Standard Model, does not meet the property of superposition. The
Lagrangian of Higgs field is
\begin{eqnarray}
 \label{eq21}
 {\cal L}_{Higgs}=-\frac{1}{2}(\partial_\mu\varphi_a)^2
-\frac{1}{2}\mu\varphi_a^2-\frac{\lambda}{4}\varphi_a^4,
\end{eqnarray}
where the last two term is the self-interaction of Higgs, which
intuitively guarantees the spontaneous symmetry breaking in vacuum
. But the last term apparently violates the rules in equation
(\ref{eq17}) now that it is nothing but a $\varphi_4$-interaction.
If Higgs exists, it shows that the spontaneous symmetry breaking
of Goldstone mechanism will terminate validity of the property of
superposition. But if the property of superposition of field is
considered a basic and universal rule in nature, Higgs will not
exist, or more precisely, the mechanism of spontaneous symmetry
breaking should be re-examined.It is the most power challenge to
the universality of the property of superposition of field of
elementary particle.

\section{Conclusions}
In summary, Considering  a field is superposable we deduce  its
dynamic equation. The general form of it is given as equation
(\ref{eq15}). Considering no higher order differentials than the
first order differential to 4-coordinates can be included in
Lagrangian, only $m=1,\ 2$ can be selected in equation
(\ref{eq15}); hence the general form of Lagrangian is given as
equation (\ref{eq17}). We examined the Lagrangians of elementary
particles and concluded that all the successful and tested
Lagrangians meet the requirement of the property of superposition.
It is verified that the property of superposition requires the
continuous transformation group a Lagrangian to be a compact Lie
group, which is consistent with $SU(N)$ gauge theory. In
particular, we argued the relation between spontaneous symmetry
breaking and the requirement of the property of superposition. It
is concluded that the existence of Higgs would violates this
property. On the contrary, if the property of superposition is
fundamental in nature, it will forbid the existence of Higgs.

{\bf Acknowledgments:} We thank  Dr J. Fu for supporting us in
many respects and their constant helps. We thank Prof. J. Li and
Z. Zhang for the earnest supervision. We thank Yanheng Li for the
fruitful discussions in this work. The work was supported in part
by the grants NSFC 10447123.

%\bibliography{clps}

\end{document}